\documentclass{aa}  

\usepackage{graphicx}
\usepackage{txfonts,booktabs}
\usepackage{hyperref}
\hypersetup{
    unicode=false,
    pdftoolbar=true,
    pdfmenubar=true,
    pdffitwindow=false,
    pdfstartview={FitH},
    pdftitle={My title},
    pdfauthor={Author},
    pdfsubject={Subject},
    pdfcreator={Creator},
    pdfproducer={Producer},
    pdfkeywords={keyword1} {key2} {key3},
    pdfnewwindow=true,
    colorlinks=true,
    linkcolor=red,
    citecolor=blue,
    filecolor=magenta,
    urlcolor=cyan
}
\usepackage{fixltx2e}
\begin{document}

   \title{An eclipsing post common-envelope system consisting of a pulsating hot subdwarf B star and a brown dwarf companion}


   \author{V. Schaffenroth\inst{1,2}, B.N. Barlow\inst{3}, H. Drechsel\inst{1}, and B.H. Dunlap\inst{4}   
          }

   \institute{Dr.\,Remeis-Observatory \& ECAP, Astronomical Institute, Friedrich-Alexander Universit\"at Erlangen-N\"urnberg, Sternwartstr.~7, 96049 Bamberg, Germany\\
\email{veronika.schaffenroth@sternwarte.uni-erlangen.de} \and Institute for Astro- and Particle Physics, University of Innsbruck, Technikerstr. 25/8, A-6020 Innsbruck, Austria \and  Department of Physics, High Point University, 833  Montlieu Ave, High Point, NC 27268  USA  \and Department of Physics and Astronomy, University of North Carolina, Chapel Hill, NC 27599-3255, USA
             }

   \date{Received 20 Jan 2015; accepted 14 Feb 2015}

 \titlerunning{A pulsating subdwarf B binary with a brown dwarf companion}
   \authorrunning{Schaffenroth et al.}

 \abstract{Hot subdwarf B stars (sdBs) are evolved, core helium-burning objects located  on the extreme horizontal branch. Their formation history is  still puzzling as the sdB progenitors must lose nearly all of their hydrogen envelope during the red-giant phase. About half of the known sdBs are in close binaries with periods from 1.2\,h to a few days, a fact that implies they experienced a common-envelope phase.  Eclipsing hot subdwarf binaries (also called HW Virginis systems) are rare but important objects for determining fundamental stellar parameters. Even more significant and uncommon are those binaries containing a pulsating sdB, as the mass can be determined independently by asteroseismology.
 Here we present a first analysis of the 
 eclipsing hot subdwarf binary V2008-1753. The light curve shows a total eclipse, a prominent reflection effect, and low--amplitude pulsations with periods from 150 to 180 s. An analysis of the light-- and radial velocity (RV) curves indicates a mass ratio close to $ q = 0.146$, an RV semi-amplitude of $K=54.6 \,\rm kms^{-1}$, and an inclination of $i=86.8^\circ$.  Combining these results with our spectroscopic determination of the surface gravity, $\log \,g = 5.83$, the best--fitting model yields an sdB mass of 0.47$M_{\rm \odot}$ and a companion mass of $69 M_{\rm Jup}$.  As the latter mass is below the hydrogen-burning limit, V2008-1753 represents the first HW Vir system known consisting of a pulsating sdB and a brown dwarf companion. Consequently, it holds great potential for better constraining models of sdB binary evolution and asteroseismology.}


   \keywords{stars: subdwarfs, binaries: eclipsing, binaries: spectroscopic, stars: brown dwarfs, stars: fundamental parameters, stars: individual: V2008-1753, stars: oscillating
               }

   \maketitle
%

\section{Introduction}
Hot subdwarf B (sdB) stars are core-helium burning stars with very thin hydrogen envelopes that are found on the extreme horizontal branch (EHB) \citep[see][for a review]{heber:2009}.  While the future evolution of sdB stars is quite certain (they will become white dwarfs), their prior evolutionary paths remain to be fully resolved.  Nonetheless, one thing seems to be assured:  the relatively high fraction of sdBs in binaries  \citep{maxted:2002,napi} implies that binary interactions play a crucial role in their formation.
\citet[see][and references therein]{han:2002,han:2003} proposed several binary--related mechanisms as formation channels for sdB stars:
\begin{itemize}

\item common envelope ejection leading to short-period binaries
with periods from 0.1-10 days with either white dwarf or low mass main sequence companions. From binary population synthesis \citep{han:2002} a mass distribution that peaks sharply around 0.47 $M_{\rm \odot}$ is expected. This mass is called the canonical mass.
\item stable Roche lobe overflow resulting in sdB binaries with masses around 0.47 $M_{\rm \odot}$ and orbital periods of 10-100 d
\item double helium white dwarf mergers giving rise to single
sdB stars with a wider distribution of masses.
\end{itemize}

Common envelope (CE) evolution is believed to be an important process for a large number and wide diversity of binary stars.   In particular, this formation process is vital to compact binary systems, as many of them  once must have been orders of magnitude
wider than their present--day separations.  Although binary population synthesis (BPS) models are, in general, able to produce compact binaries through common envelope evolution, this  important phase in binary evolution is still poorly understood \citep[see][for a review]{ivanova}.  Many of the CE--related parameters in BPS codes remain unconstrained by observations \citep{clausen:2012}, including common--envelope ejection efficiency, minimum mass for core He ignition, and  envelope binding energy.  Consequently, the investigation of post--CE systems, especially those containing pulsators or exhibiting eclipses, can improve our understanding of this highly--important stage of stellar evolution.

Some have proposed that a planet or brown dwarf companion might also be able to trigger the loss of envelope mass in the red-giant phase of the sdB progenitor \citep{soker,nt,han:2012}. When the host star evolves and becomes a red giant, close substellar companions will be engulfed in a common envelope. The outcome of this interaction is unclear, but there are three distinct possibilities: the substellar object can either survive, merge with the core of the red giant, or evaporate. The latter two outcomes might provide an explanation for some of the observed single sdBs. Discoveries of brown dwarfs in close, eclipsing sdB systems \citep{geier,vs:2014_I} support the idea that substellar companions are sufficient for ejecting the envelopes of red giants. This scenario is supported also by the discovery of a brown dwarf ($M_{\rm BD} = 0.053 \pm 0.006\,M_{\rm \odot}$) in a close orbit (0.08\,d) around a low-mass white dwarf \citep{nature:maxted}. 
Consequently, substellar companions are able to influence late stellar evolution.

Eclipsing binaries are of special interest as they allow the determination of the masses and radii of both components as well as the period and separation of the system. The mass in particular is vital to constraining sdB formation models. Eclipsing sdB binaries with low mass main sequence or brown dwarf companions (HW Vir systems) are easily recognizable by the shapes of their light curves, which are dominated by both eclipses and a strong reflection effect due to the large temperature difference between both components. The amplitude of the reflection effect depends on the separation distance, the temperature of the sdB, and the albedo of the companion.  In addition to shedding light on sdB evolution, studies of these relatively rare, post--common envelope systems can be used to constrain current models of common-envelope evolution \citep[e.g.][for eclipsing WD binaries]{zorotovic:2010}. 

Additional information concerning the structure and evolution of sdBs may be found by studying their pulsations.  The $\rm sdBV_r$ stars, discovered by \citet{kilkenny:1997} and independently theoretically predicted by \citet{charpinet:1996}, are low amplitude, multimode pulsators with typical periods ranging between 80-600\,s. Their pulsation amplitudes are generally of the order of a few mmag. The short periods, being of the order of and shorter than the radial fundamental mode for these stars, suggest that the observed modes are low-order, low-degree p-modes \citep{charpinet:2000}. The known $\rm sdBV_r$ stars occupy a region in the $T_{\rm eff}-\log{g}$ plane with effective temperatures between 28\,000~K and 36\,000~K and surface gravities (log~g) between 5.2 and 6.2. Only 10\,\% of all stars falling in this region show pulsations. Moreover, also sdBs with slow high radial-order g-mode oscillations with periods on the order of $2000-8000$~s and termperatures of $22000-29000$~K were found \citep{green:2003}. In the overlap region also hybrid pulsators showing p-mode as well as g-mode pulsations exist \citep{schuh:2006}. Asteroseismology of these stars allows for an independent and accurate determination of the sdB mass. 

The problem with typical HW Vir systems has been that they are single--lined spectroscopic binaries \citep[see e.g.][]{vs,vs:2014_I}, and as such it is normally not possible to derive a unique mass ratio.   HW Vir systems harboring a {\em pulsating} sdB primary, however, offer additional possibilities, as the stellar properties can be constrained by the light curve and asteroseismological analyses.   Until recently, only two HW~Vir systems with pulsating sdBs were known. The first such object -- NY~Vir -- was found to be a pulsating sdB in an eclipsing binary by \citet{kilkenny:1998}. It shows more than 20 pulsation modes with amplitudes of several mmag.An astroseismic analysis was able to determine the stellar parameters of this system independently from the lightcurve analysis \citep{vangrootel}. \citet{oestensen:2010} found another pulsating sdB\,+\,dM HW~Vir binary (2M1938+4603). Unfortunately, the amplitudes of the pulsations, which were detected by {\em Kepler} are so small that they cannot be observed by ground-based telescopes. Thus, this star is therefore not an ideal target for asteroseismological modelling.

\citet{van_noord} reported the discovery of a promising new HW~Virginis system, V2008-1753 (CV=16.8 mag), which was found during an automatic search for variable stars conducted with the 0.4\,m Calvin College Robotic Telescope in Rehoboth, New Mexico. Their relatively noisy light curve showed eclipses and a strong reflection effect.  Interestingly, this sdB binary has an orbital period of only 1.58\,h, the shortest period ever found in a HW~Virginis system.  Here we present the first thorough analysis of this unique system, along with the discovery of low--amplitude pulsations in the sdB primary. Section \ref{observations} describes the observational data. The analysis is explained in Sects. \ref{spec} (spectroscopy) and \ref{photo} (photometry). Evidence for the brown dwarf nature of the companion is discussed in Sect. \ref{nature}. Finally, we end up with conclusions and suggest further opportunities that are offered by this one-of-a-kind system.


\section{Observations}
\label{observations}
\subsection{Time--Series Spectroscopy}

We used the Goodman Spectrograph on the 4.1--m SOuthern Astrophysical Research (SOAR) telescope to obtain time--series spectroscopy of V2008-1753 over a full orbital cycle and determine the orbital velocity of its primary sdB component.   A 1.35'' longslit and a 930 mm$^{-1}$ VPH grating from Syzygy Optics, LLC, were employed to cover the spectral range 3600--5250 \AA\ with an approximate resolution of 3.8 \AA\ (0.84 \AA\  per binned pixel).  The position angle was set to 197.8 degrees E of N so that we could place a nearby comparison star on the slit, with the intention of characterizing and removing instrumental flexure effects.  In order to maximize our duty cycle, we binned the spectral images by two in both the spatial and dispersion directions and read out only a 2071 x 550 (binned pixels) subsection of the chip.  Each exposure had an integration time of 90 s, with 8 s of overhead between successive images, yielding a duty cycle of approximately 92\%.   Overall, we acquired 70 spectra of V2008-1753 between 23:49:20.87 UT (2013-08-31) and 01:42:07.17 UT (2013-09-01).  The airmass decreased from 1.22 to 1.03 over this time period.  Upon the completion of the time series, we obtained several FeAr comparison lamp spectra and quartz--lamp spectra for wavelength calibration and flat--fielding.

Standard routines in {\sc IRAF}, primarily {\em ccdproc}, were used to bias--substract and flat--field the spectral images.  Given the low thermal noise in the spectrograph system, we did not subtract any dark frames as we wanted to avoid adding noise to the images.  We used {\em apall} to optimally extract one--dimensional spectra and subtract a fit to the sky background for both the target and constant comparison star.  Finally, we wavelength--calibrated each spectrum using the master FeAr comparison lamp spectrum taken at the end of the time--series run.  The resulting individual spectra of V2008-1753 had a signal--to--noise ratio (S/N) around 15--20 pixel$^{-1}$, while the individual spectra of the comparison star (which looked to be G or K--type), had a S/N twice as high.

\subsection{Time--Series Photometry}
High--precision photometry was acquired with SOAR/Goodman through i' and g' filters on 15/16 August 2013.  Although our primary goal was to model the binary light curve, our secondary goal was to look for stellar pulsations, and thus we used an instrumental setup that had both a high duty cycle and a Nyquist frequency above those of most known sdB pulsation modes.   We binned the images 2 x 2 and read out only a small 410 x 540 (binned pixel) subsection of the chip that included both the target and 10 comparisons stars with the same approximate brightness level.  We achieved an 87\% duty cycle with 25-s exposures for the g' light curve, and a 81.4\% duty cycle with 14-s exposures for the i' light curve.  In both cases, we observed the field for at least one full orbital cycle near an airmass of 1.1.   We concluded each night with a set of bias frames and dome flats.  Thermal count rates were too low to warrant the acquisition of dark frames.

All object images were bias--subtracted and flat--fielded using IRAF's {\em ccdproc} package.  We extracted light curves with an IDL program we wrote that uses the function APER, which is based on DAOPHOT \citep{stetson:1987}.  We produced light curves over a wide range of aperture radii and selected the aperture that maximized the S/N in the light curve.  To mitigate the effects of atmospheric extinction and transparency variations, we divided the light curve of V2008-1753 by the average of those of the constant comparison stars, after verifying they were indeed non--variable.  
Residual extinction effects are often removed in light curves of this duration by fitting and dividing the light curve by the best-fitting parabola.  However, given the large-amplitude binary signals present, this process would distort the actual stellar signal,  Instead, we fit straight lines through points of the same phase and, informed by these fits, removed the overall 'tilt' in each light curve.  While not perfect, this process helps to mollify the effects of residual extinction variations.  We then divided each light curve by its mean value and subtracted a value of one from all points to put them in terms of fractional amplitude variations.


\section{Spectroscopic analysis}
\label{spec}
\subsection{Radial velocity curve}
\begin{figure}
\centering
\includegraphics[width=1.0\linewidth]{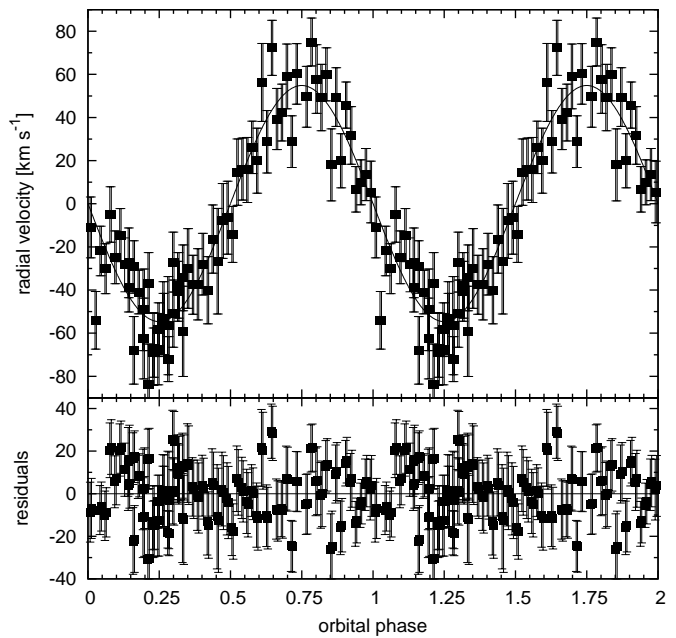}
\caption{{\em Top panel:} Radial velocity curve for the sdB primary in V2008-1753, plotted twice for better visualization.  The solid line denotes the best--fitting circular orbit to the data. {\em Bottom panel:}  Residuals after subtracting the best--fitting sine wave from the data.}
\label{rv}
\end{figure}

Radial velocity shifts were determined by measuring the positions of the hydrogen Balmer profiles H$\beta$ through H9; although H10, H11, and several He I lines were also present, they were too noisy to provide reliable positions.  We used the MPFIT routine in IDL \citep{markwardt:2009}, which relies on the Levenberg-Marquardt method, to fit simple inverse Gaussians to the line profiles and determine their centroids.  The only available guide star near our field was significantly redder than the sdB target, and, consequently, its use led to a gradual shift in the slit alignment over the course of our observations (Goodman had no atmospheric dispersion corrector at the time); this drifting results in a colour--dependent velocity shift.  Additionally, instrumental flexure as the Nasmyth cage rotates also affected the stars' alignment on the slit, although only slightly.  We removed both of these time--dependent wavelength solution effects by tracking the absorption--line features in the constant comparison star.  Figure \ref{rv} presents the resulting radial velocity curve for V2008-1753, plotted twice for better visualisation.

We again used MPFIT to fit a sine wave to the data and determine the semi--amplitude of the velocity variation; the orbital period and phase were fixed during this process.  We derive an orbital velocity of $K = 54.6 \pm 2.4 \,\,\mathrm{km  \, s^{-1}}$ for the sdB primary.  Eccentric orbits were also fitted to the radial velocity curve, but as we currently have no reason to prefer them over $e=0$, we continue the analysis under the assumption that the orbit is circular. Residuals from the best--fitting sine wave are shown in the bottom panel of Figure \ref{rv} and are consistent with noise.  The mean noise level in the Fourier transform of the residual velocity curve is $2 \, \mathrm{km \, s^{-1}}$.

\subsection{Atmospheric parameters}
\begin{figure} 
\centering
\includegraphics[angle=-90,width=0.8\linewidth]{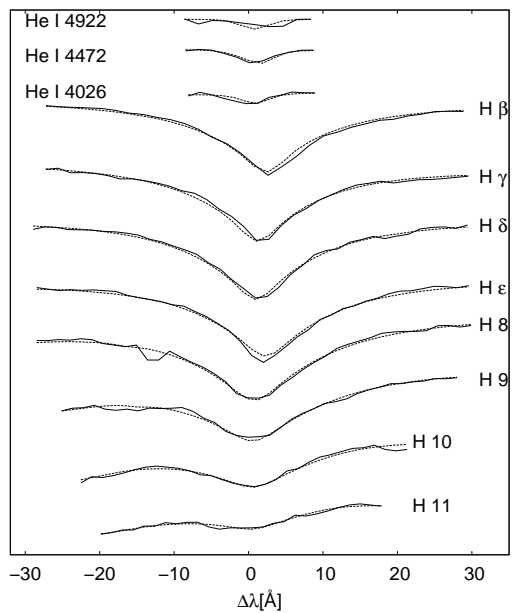}
\caption{ Fit of the Balmer and helium lines in the co-added SOAR spectrum. The solid line shows the measurement, and the dashed line shows the best fitting synthetic spectrum. As the H $\epsilon$ seems to be effected by a blend of the Ca line next to it, it was excluded from the fit.
}
\label{linienplot}
\end{figure}

In preparation for determining the atmospheric parameters of the sdB, we first de--shifted all individual spectra according to our orbital solution above and then co--added them to improve the overall S/N. We fit synthetic spectra, which were calculated using local thermodynamical equilibrium model atmospheres with solar metallicity and metal line blanketing \citep{heber:2000}, to the Balmer and helium lines of the co-added SOAR spectrum using SPAS \citep[SPectral Analysis Software,][]{hirsch}.  The best--fitting synthetic spectrum had the following orbital parameters:
\begin{align*}
T_{\rm eff}&=32800\pm 250\\
\log{g}&=5.83\pm0.04\\
\log{y}&=-2.27 \pm 0.13
\end{align*}
with 1-$\sigma$ statistical errors determined by bootstrapping. 
For some sdB systems with reflection effects, an analysis of spectra with sufficiently high signal-to-noise taken at different phases shows that the atmospheric parameters apparently vary with phase \citep[e.g.][]{vs,vs:2014_I}. These variations can be explained by the companion's contribution to the spectrum (reflection only) varying with orbital phase. Systems with similar parameters show such variations in temperature and surface gravity on the order of $1000 - 1500 \rm\,K$ and 0.1 dex, respectively.  To account for the apparent change in the parameters we formally adopt the values determined from the co-added spectrum, which represents a mean value, with a larger error: $T_{\rm eff}=32800\pm 750$ K and $\log{g}=5.83\pm0.05$ for the sdB. Figure \ref{linienplot} shows the corresponding fit of the Balmer and helium lines. We excluded H$\epsilon$ from the fit, as this line is mostly blended with the Ca~II H-line and hence is less well represented by this fit.

\begin{figure}
\centering
\includegraphics[width=1.0\linewidth]{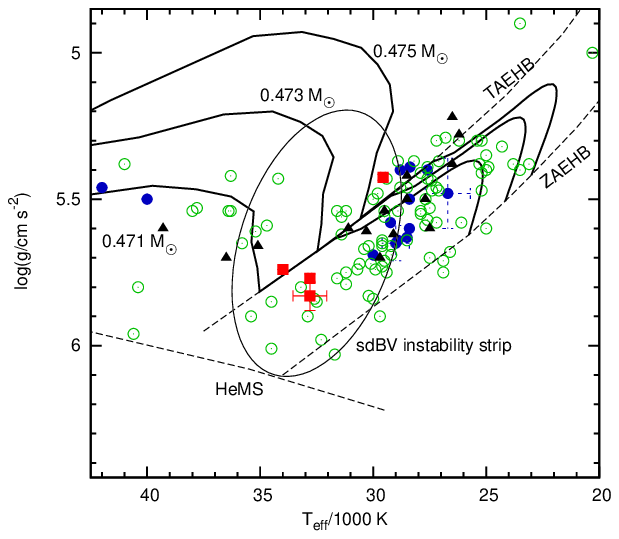}
\caption{$\rm T_{eff}-\log{g}$ diagram of the HW~Vir systems. The helium main sequence (HeMS) and the EHB band (limited by the zero-age EHB, ZAEHB, and the terminal-age EHB, TAEHB) are superimposed by evolutionary tracks by \citet{Dorman:1993} for sdB masses of 0.471, 0.473, and 0.475\,$M_{\odot}$. The positions of the HW~Vir systems with pulsating sdBs -- V2008-1753 (this work, with error bars), NY Vir \citep{vangrootel}, 2M1938+4603 \citep{oestensen:2010}, and PTF1 J072456+125301 (Schindewolf et al. submitted, Kupfer priv. com.) -- are marked by red squares. Blue dots mark the positions of other HW~Vir-like systems \citep[][Schaffenroth et al., in prep, Kupfer, priv. com.]{drechsel:2001, for:2010, maxted:2002, klepp:2011, oestenson:2008, Wood:1999,almeida:2012,barlow:2012,vs, tuc_schaff}. The positions of the two HW~Vir systems with BD companions J1622 \citep{vs:2014_I} and J0820 \citep{geier} are indicated by the blue dots with error bars. The black triangles mark sdB+dM systems showing a reflection effect but no eclipses \citep[and references therein]{kupfer:2015}. The green, open dots represent other sdB+WD binaries or sdB binaries with unknown companion type \citep{kupfer:2015}. The approximate location of the $\rm sdBV_r$ instability strip is indicated by an ellipse.}

\label{teff-logg}
\end{figure}

The $T_{\rm eff}-\log{g}$ diagram is displayed in Fig. \ref{teff-logg} and shows that V2008-1753 lies in the middle of the extreme horizontal branch.  Although it was previously suggested that HW~Virginis systems cluster together only in a small part of the $T_{\rm eff}-\log{g}$ diagram \citep{vs:2014_I}, the position of V2008--1753 seems to go against this hypothesis. However, it is still apparent that most of the known HW Vir systems and reflection effect binaries (both sdB+dM systems with different inclinations) concentrate in a distinct region between a $T_{\rm eff}$ of 26000-30000 K and a $\log{g}$ of about 5.3 to 5.7, only at the edge of the instability strip. There are five exceptions of binaries with sdOB star primaries and M star companions at higher temperatures, which are possibly just more evolved. Yet, the three HW~Vir systems showing short-period p-mode pulsations with amplitudes observable from the ground lie in a different part of the $T_{\rm eff}-\log{g}$ diagram, nearer to the He-MS, in the central part of the instability strip for $\rm sdBV_r$s, as expected. In contrast to that the other sdB binaries with either white dwarf secondaries or companions of unknown type do not show any clustering but are uniformly distributed over the EHB. This could indicate that the sdBs with low-mass main sequence companions differ from the sdBs with white dwarf companions.
\section{Photometric analysis}
\label{photo}
\subsection{Pulsations}
\label{pulsa}
\begin{figure}
\centering
\includegraphics[width=1.0\linewidth]{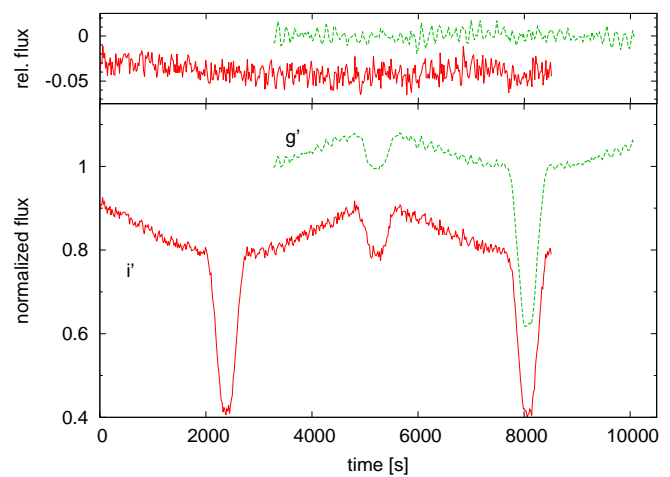}
\caption{g' (dashed line) and i' (solid line) filter light curves of V2008-1753 taken with SOAR. The two light curves were taken
in subsequent orbital cycles. The sub-figure shows the pulsation signal after the subtraction of the binary signal by the best-fitting light curve model.}
\label{pulsation}
\end{figure}

Both the g' and i' light curves from the SOAR telescope exhibit pulsations too low in amplitude to have been detected in the data analyzed by \citet{van_noord}. In order to disentangle the pulsations from the binary effects in the light curve, we used an approach similar to that demonstrated by \citet{vuckovic:2007}. First, we fit the eclipses and reflection effect as described in Section \ref{ecl}) and subtracted the best-fitting model from the observed light curve. The original g' and i' filter light curves are displayed in Fig. \ref{pulsation}, along with the same curves after the subtraction of the binary signal.  The pulsations are visible by eye in the g' curve but less apparent in the i' data, due to its lower S/N and lower pulsation amplitudes at redder wavelengths. The large--scale trends in both residual light curves are likely due to residual atmospheric extinction and transparency variations.

\begin{table}
\caption{Pulsation frequencies and amplitudes}
\label{puls}
\centering
\begin{tabular}{ccccc}
		\toprule
	   &f\tablefootmark{a,}\tablefootmark{b}&amplitude&phase\tablefootmark{c}&S/N\\
	   &[mHz]&[ppt]&&\\
	   		\toprule
	     F1   & 6.565$\pm$0.005 &   3.5$\pm$ 0.2 & 0.739$\pm$0.019 & 14.5 \\
	     F2   & 5.494$\pm$0.008 &   3.1$\pm$ 0.2 & 0.82$\pm$0.03 & 12.6 \\
	     F3   & 6.289$\pm$0.006 &   3.1$\pm$ 0.2 & 0.74$\pm$0.02 & 12.5 \\
	     F4   & 5.638$\pm$0.011 &   2.2$\pm$ 0.2 & 0.71$\pm$0.05 & 9.2 \\  
	     \toprule
	     F1   &  6.572$\pm$0.006 &   3.2$\pm$ 0.3&0.12$\pm$0.03 & 8.7\\ 
	     F2   & -- & -- & -- & --\\
	     F3   &  6.295$\pm$0.006 &   2.8$\pm$ 0.3&0.90$\pm$0.04 & 7.6\\
	     F4   &  5.685$\pm$0.008 &   2.3$\pm$ 0.3&0.87$\pm$0.03 & 6.2\\
	    \bottomrule
	    
\end{tabular}
\tablefoot{
\tablefoottext{a}{upper half: frequencies found in g' light curve\\
				 lower half: frequencies found in i' light curve	}\\

\tablefoottext{b}{errors as given by FAMIAS, more realistic is an error around 0.1-0.2 mHz}\\
\tablefoottext{c}{reference time: first point of lightcurve\vspace{-0.3cm}
\begin{itemize}
\item g': $\rm BJD_{\rm TBD}$ = 2456519.5846905
\item i': $\rm BJD_{\rm TBD}$ = 2456520.53415896
\end{itemize}	}			 
}
\end{table}

\begin{figure*}
\centering
\includegraphics[angle=-90,width=0.49\linewidth]{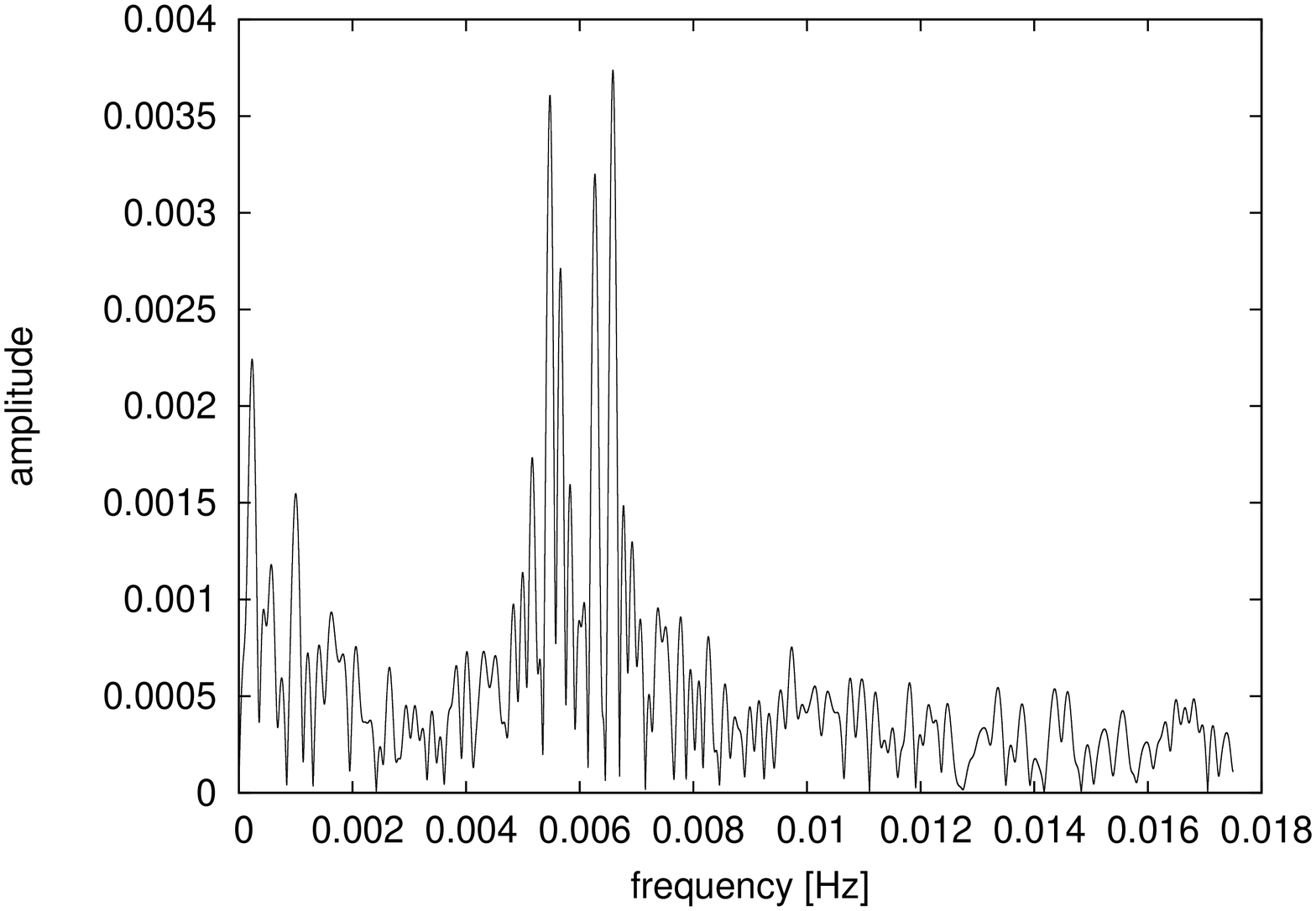}
\includegraphics[angle=-90,width=0.49\linewidth]{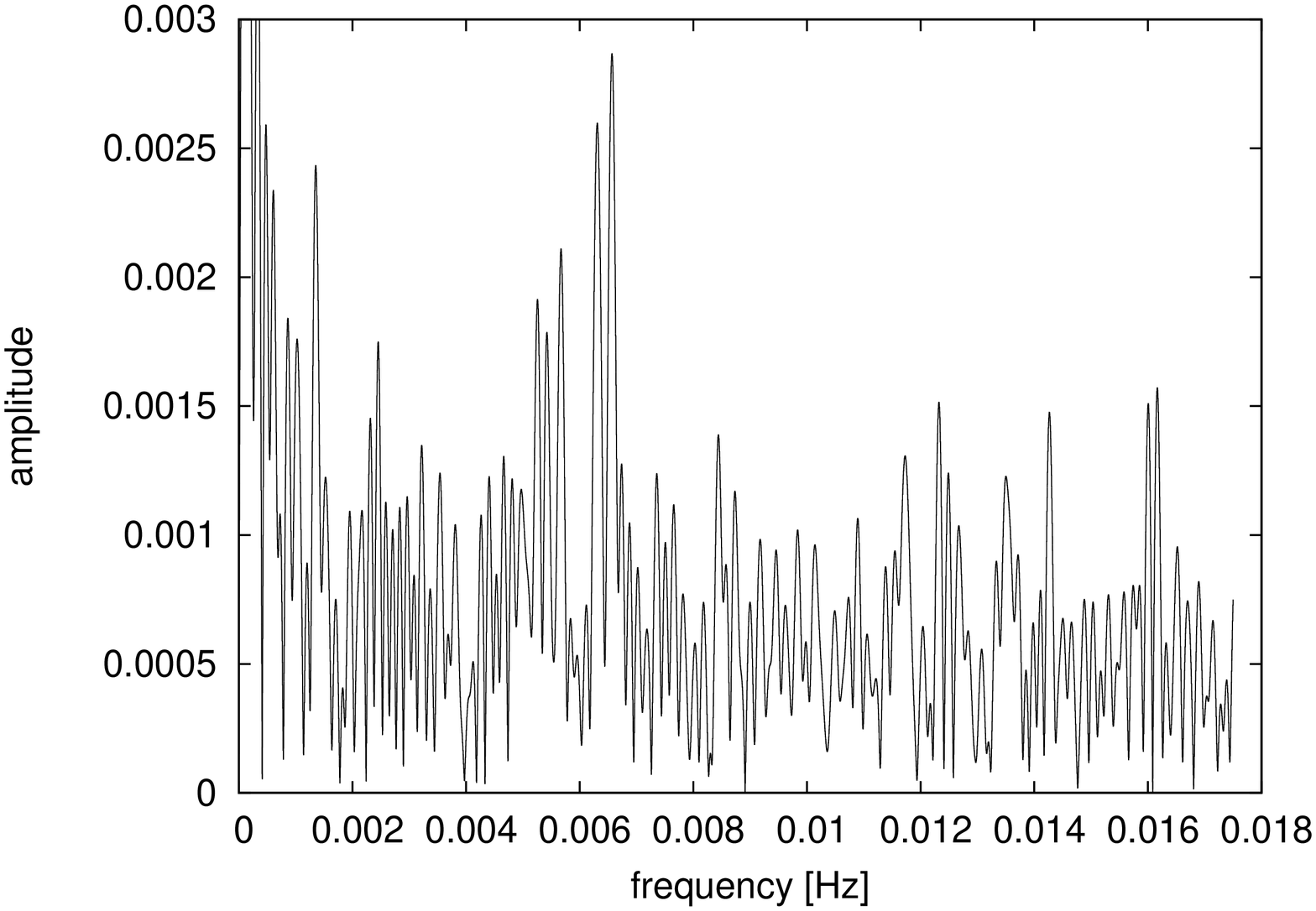}
\caption{Fourier transform of the pulsation signal shown in the sub-figure of Fig. \ref{pulsation}. The left and right figures show the FTs of the g' and the i' filter light curves, respectively.}
\label{ft}
\end{figure*}

We calculated the Fourier transformations (FTs) of the light curves using FAMIAS\footnote{\url{http://www.ster.kuleuven.be/~zima/famias}} \citep{zima}. The resulting FTs are  displayed in Fig. \ref{ft}.  We detect at least four independent pulsation modes with periods ranging from 2.5 to 3 min and amplitudes $<$ 4 ppt. The best--fit frequencies, along with their amplitudes and S/N, are listed in Table \ref{puls}.  The mode with the second--highest amplitude in the g' light curve (F2) was not clearly detected in the i' light curve, but its apparent absence might be explained by the high noise level in this data set.   The elevated power at lower frequencies is likely due to inaccuracies in the binary light curve modeling, along with the atmospheric extinction and transparency corrections.  A much longer time base is needed to improve the frequency determination and to be able to use the pulsations for asteroseismology.  Consequently, we do not perform a more thorough pulsation analysis than this and limit our result simply to the detection of pulsations alone.  To prepare the light curve for binary modeling, we use the results from Table \ref{puls} to subtract the detected pulsations from the original light curves.  

\subsection{Binary Light Curve Modeling}  
\label{ecl}
The binary light curve exhibits all the typical features of an HW~Vir system. Due to the short period and the relatively high temperature of the subdwarf, the reflection effect is rather strong, with an amplitude around 10\,\%.  The secondary eclipse appears to be nearly total, in accordance with the high inclination and the very deep primary eclipse. As our light curves only cover one complete orbital cycle, it is not possible to determine an accurate period from our data alone. For the ephemeris we hence cite the period derived by \citet{van_noord}, which was determined from a much larger baseline.  We were able to determine precise eclipse timings using our data by fitting parabolas to the minima. They are summarized in Table \ref{eclipse}.  Using these values and the period from \citet{van_noord}, the ephemeris of the primary minimum is given by
\begin{equation}
\rm BJD_{\rm TBD} =  2456519.64027(1)+ 0.065817833(83) \cdot E
\end{equation}
where $E$ is the cycle number.  The error in the period quoted by \citet{van_noord} is likely to be an underestimate given the omission of systematics and the poor sampling.  We believe an error of 0.0001 to be more appropriate.  A comparison of the secondary eclipse compared to the primary eclipse of both lightcurves separately reveals a slight departure of the secondary mid-eclipse from phase 0.5; such an offset can be caused by both the R\o mer delay (extra light travel time due to the binary orbit) and an eccentricity $e>0$.  For small eccentricities, the total shift of the secondary eclipse with respect to phase 0.5 is given in \citet{barlow:romer}:
\begin{equation*}
\Delta t_{\rm SE}\backsimeq\Delta t_{\text{R\o mer}}+\Delta t_{\rm ecc}\backsimeq\frac{PK_{\rm sdB}}{\pi c}\left(\frac{1}{q}-1\right)+\frac{2 P}{\pi}\,e\cos{\omega}
\end{equation*}
From both light curves we measure a shift of $3\pm1$ s between the time of the secondary minimum and phase 0.5. With our system parameters we would expect a theoretical shift of the secondary eclipse with respect to phase 0.5 due to the R\o mer delay of 2~s. If we take that into account, we would get a maximal eccentricity of $e\cdot\cos{\omega} < 0.00055$. Ignoring the R\o mer delay results in a maximum eccentricity of $e\cdot\cos{\omega} < 0.0011$. Because of the large errors in the radial velocity determinations, the eccentricity cannot be constrained by the radial velocity curve to that precision.

\begin{table}
\caption{Eclipse times}
\label{eclipse}
\centering
\begin{tabular}{ccc}
\toprule
filter&eclipse&$\rm BJD_{\rm TBD}$[d]\\
\toprule
g'&primary&2456519.64027(1)\\
g'&secondary&2456519.60733(2)\\
i'&primary&2456520.56176(1)\\
i'&secondary&2456520.59463(1)\\
i'&primary&2456520.62757(1)\\
\bottomrule
\end{tabular}
\end{table}

\begin{figure}
\centering
\includegraphics[width=1.0\linewidth]{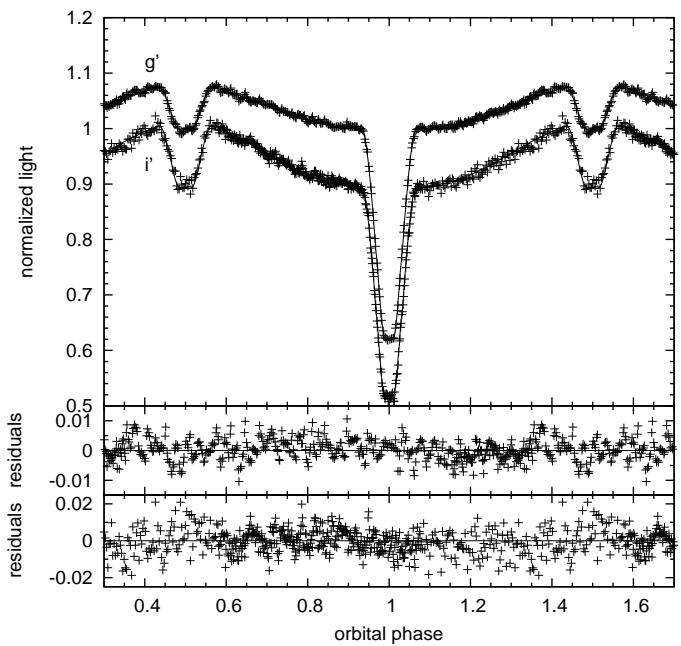}
\caption{g' and i' light curves after the removal of the pulsation signal as explained in Sect. \ref{pulsa} together with the best-fitting light curve model. The residuals displayed in the two lower panels still show signs of low-amplitude pulsations.}
\label{lc_model}
\end{figure}

A photometric solution to the binary light curve (with pulsations removed; Fig. \ref{lc_model}) was determined using MORO \citep[MOdified ROche program, see][]{drechsel:1995}. This program calculates synthetic light curves which we fit to the observations using the SIMPLEX algorithm. This light curve solution code is based on the Wilson-Devinney approach \citep{wilson:1971} but uses a modified Roche model that considers the mutual irradiation of hot components in close binary systems. More details of the analysis method are described in \citet{vs}.
To calculate the synthetic light curves, 12\,+\,5n (n is the number of light curves) parameters are used. Such a high number of partially--correlated parameters will inevitably cause severe problems if too many and wrong combinations are adjusted simultaneously. In particular, there is a strong degeneracy with respect to the mass ratio. After the orbital inclination, this parameter has the strongest effect on the light curve, and it is highly correlated with the component radii. Hence we kept the mass ratio fixed at certain values and calculated solutions for these mass ratios, which were subsequently compared and evaluated according to criteria explained below. 

Given the large number of parameters present in the code, it is imperative to constrain as many parameters as possible based on independent inputs from spectroscopic analyses or theoretical constraints.  From the spectroscopic analysis, we derived the effective temperature and the surface gravity of the sdB primary and fixed these parameters during the fitting. Due to the early spectral type of the primary star, the gravity darkening exponent was fixed at $g_1=1$, as expected for radiative outer envelopes \citep{zeipel:1924}. For the cool convective companion, $g_2$ was set to 0.32 \citep{lucy:1967}. The linear limb darkening coefficients were extrapolated from the table of \citet{claret}.

To determine the quality of the light curve fit, the sum of squared residuals $\sigma$ of all observational points with respect to the synthetic curve was calculated as a measure of the goodness of the fit. Unfortunately, the $\sigma$ values of the best light curve fits for the different mass-ratios did not differ significantly. Therefore, we cannot determine a unique solution from the light curve analysis alone. 
The full set of parameters describing the best--fitting solution for a mass ratio of $q=0.146$, which corresponds to an sdB with the canonical mass of 0.47 $M_\sun$ are given in Table \ref{asas}, with errors determined by the bootstrapping method. The light curves in the $g'$ and $i'$ bands are displayed in Fig.~\ref{lc_model} together with the best-fit models for these parameters. The parameters of the system derived from this lightcurve solution together with the semi-amplitude of the RV curve are summarized in Table \ref{mass}. The errors result from error propagation of the errors in $K$ and $P$.
\begin{table}
\caption{ Adopted light curve solution.}
\label{asas}
\begin{tabular}{lcl}
\hline
\noalign{\smallskip}
\multicolumn{3}{l}{Fixed parameters:}\\
\noalign{\smallskip}
\hline
\noalign{\smallskip}
$q\,(=M_{2}/M_{1})$ & & $0.146$\\
$T_{\rm eff}(1)$&[K]&\multicolumn{1}{l}{33000}\\
$g_1^b$&&\multicolumn{1}{l}{1.0}\\
$g_2^b$&&\multicolumn{1}{l}{0.32}\\
$x_1(g')^c$&&\multicolumn{1}{l}{0.21}\\
$x_1(i')^c$&&\multicolumn{1}{l}{0.14}\\
$\delta_2^d$&&\multicolumn{1}{l}{0.0}\\
\noalign{\smallskip}
\hline
\noalign{\smallskip}
\multicolumn{3}{l}{Adjusted parameters:}\\
\noalign{\smallskip}
\hline
\noalign{\smallskip}
$i$ & [$^{\rm \circ}$] & $86.83 \pm 0.45$\\
$T_{\rm eff}(2)$ & [K]& $2960 \pm 550$\\
$A_1^a$&&\multicolumn{1}{l}{$1.0\pm0.002$}\\
$A_2^a$ & & $1.2 \pm 0.05$\\
$\Omega_1^f$&&$4.10 \pm 0.05$\\
$\Omega_2^f$&&$2.389 \pm 0.008$\\
$\frac{L_1}{L_1+L_2}(g')^g$&&$0.99995 \pm 0.00007 $\\
$\frac{L_1}{L_1+L_2}(i')^g$&&$0.99926 \pm 0.00068 $\\
$\delta_1$&&$0.026 \pm 0.01$\\
$x_2(g')$&&$0.44 \pm 0.06$\\
$x_2(i')$&&$0.62 \pm 0.07$\\
$l_3(g')^f$&&\multicolumn{1}{l}{$0.007 \pm 0.001$}\\
$l_3(i')^f$&&\multicolumn{1}{l}{$0.0 \pm 0.0$}\\
\noalign{\smallskip}
\hline
\noalign{\smallskip}
\multicolumn{3}{l}{Roche radii$^h$:}\\
\noalign{\smallskip}
\hline
\noalign{\smallskip}
$r_1$(pole)&[a]&$0.246 \pm 0.001 $\\
$r_1$(point)&[a]&$0.249 \pm 0.002 $\\
$r_1$(side)&[a]&$0.249 \pm 0.002 $\\
$r_1$(back)&[a]&$0.249 \pm 0.002 $\\
\noalign{\smallskip}
$r_2$(pole)&[a]&$0.150 \pm 0.001$\\
$r_2$(point)&[a]&$0.154 \pm 0.002 $\\
$r_2$(side)&[a]&$0.152 \pm 0.001 $\\
$r_2$(back)&[a]&$0.158 \pm 0.001 $\\
\noalign{\smallskip}
\hline
\end{tabular}\\
\tablefoot{\\
$^{a}$ Bolometric albedo\\
$^{b}$ Gravitational darkening exponent\\
$^{c}$ Linear limb darkening coefficient; taken from \citet{claret} \\
$^{d}$ Radiation pressure parameter, see \citet{drechsel:1995}\\
$^{e}$ Fraction of third light at maximum\\
$^{f}$ Roche potentials\\
$^{g}$ Relative luminosity; $L_2$ is not independently adjusted, but recomputed from $r_2$ and $T_{\rm eff}$(2)\\
$^{h}$ Fractional Roche radii in units of separation of mass centres}
\end{table}

\begin{table}
\caption{Parameters of V2008-1753}
\label{mass}
\centering
\begin{tabular}{c|c|c}
		\toprule
		\multicolumn{3}{c}{V2008-1753}\\ 
		coordinates	&\multicolumn{2}{c}{20 08 16.355\,\,-17 53 10.52  (J2000.0)}\\
		$\rm cv$&[mag]&16.8\\
		\toprule
		$i$&$^\circ$&$86.83\pm0.45$\\
		$K$&[$\rm km\,s^{-1}$]&$54.6\pm2.4$\\
		$P$&[h]&$1.5796280\pm0.0000002$\\
		 $M_{\rm sdB}$ & [$M_{\rm \sun}$] & $0.47\pm0.03$\\
		 $M_{\rm comp}$ & [$M_{\rm \sun}$] & $0.069\pm0.005$\\
		 $a$ & [$R_{\rm \sun}$] & $0.56\pm 0.02$\\
		$R_{\rm sdB}$ & [$R_{\rm \sun}$]& $0.138\pm 0.006$\\
		$R_{\rm comp}$ & [$R_{\rm \sun}$]&$0.086\pm 0.004$\\
		$\log{g}(\rm sdB,phot)$ & & $5.83\pm0.02$\\
	    $\log{g}(\rm sdB,spec)$ & & $5.83\pm0.05$\\
	    $ T_{\rm eff,sdB}$ & [K] & $32800\pm750$\\
	    \bottomrule
\end{tabular}
\end{table}

\section{The brown dwarf nature of the companion}
\label{nature}
\begin{figure}
\centering
\includegraphics[angle=-90,width=1.0\linewidth]{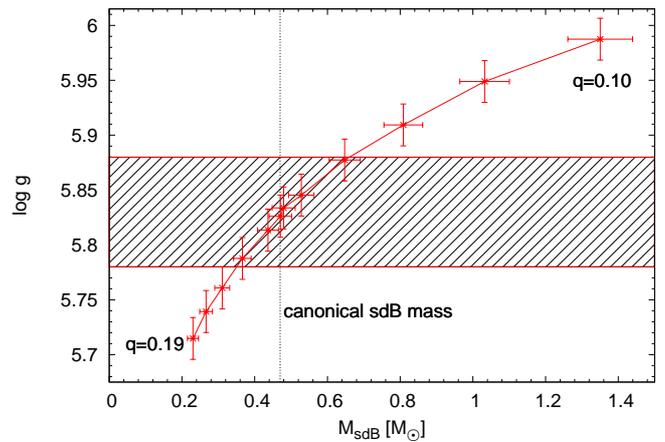}
\caption{Comparison of the photometric and spectroscopic surface gravity for the solutions with different mass ratio $q=$ 0.10, 0.11, 0.12, 0.13, 0.14, 0.145, 0.146, 0.15, 0.16, 0.17, 0.18, 0.19 (marked by the error cross). The spectroscopic surface gravity with uncertainty is given by the shaded area.}
\label{sdB}
\end{figure}

From the semi-amplitude of the radial velocity curve, the orbital period, and the inclination, we can derive the masses and radii of both components for each mass ratio.
The masses follow from:
\begin{align*}
\centering
M_1&=\frac{PK_1^3}{2\pi G}\frac{}{}\frac{(q+1)^2}{(q\cdot\sin{i)}^3}\\
M_2&=q\cdot M_1\\
\end{align*}
and the fractional radii of the light curve solution together with 
\begin{align*}
a&=\frac{P}{2\pi}\frac{K_1}{\sin{i}}\cdot\left(\frac{1}{q}+1\right)\\
\end{align*}
yield the radii. For each mass ratio we get different masses and radii. It was stated already in Sect. \ref{ecl} that it is not possible to determine the mass ratio from the light curve analysis alone. However, from the sdB mass and radius determined by the light curve analysis we can calculate a photometric surface gravity and compare that to the surface gravity derived by the spectroscopic analysis. The result is shown in Fig. \ref{sdB}. An agreement of spectroscopic and photometric surface gravity values is reached for solutions that result in sdB masses between 0.35 and 0.62\,$M_{\rm\odot}$. It is therefore possible to find a self-consistent solution. This is a fortunate situation, because gravity derived from photometry was found to be inconsistent with the spectroscopic result in other cases, such as AA~Dor \citep{aador}.

\begin{figure}
\centering
\includegraphics[width=1.0\linewidth]{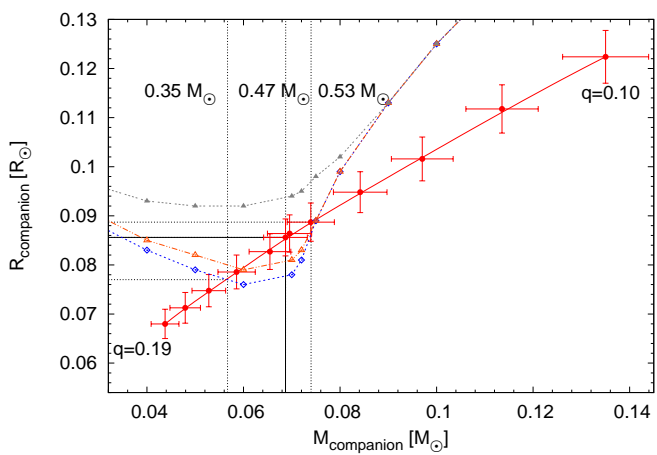}
\caption{Comparison of theoretical mass-radius relations of brown dwarfs by \citet{baraffe} for an age of 1~Gyr (filled triangles), 5~Gyrs (triangles) and 10~Gyrs (diamond) to results from the lightcure analysis. Each error cross represents a solution from the light curve analysis for a different mass ratio ($q=$ 0.10, 0.11, 0.12, 0.13, 0.14, 0.145, 0.146, 0.15, 0.16, 0.17, 0.18, 0.19). The dashed vertical lines mark different values of the corresponding sdB masses. The solid line marks the most probable solution with $q = 0.146$, which results in an sdB mass of 0.47\,$M_{\rm\odot}$.}
\label{bd}
\end{figure}

To constrain the solutions even more, we can also use theoretical mass-radius relations for the low-mass companions by \citet{baraffe} and compare them to the masses and radii of the companion derived by the light curve solutions for the various mass ratios. This was done in a similar way as in \citet{vs:2014_I}. This comparison is displayed in Fig. \ref{bd}. Relations for different ages of 1, 5 and 10~Gyrs were used. The measured mass-radius relation is well matched by theoretical predictions for stars $\gtrsim 3$ Gyrs for companion masses between 0.056\,$M_{\rm \odot}$ and 0.073\,$M_{\rm \odot}$. The corresponding mass range for the sdB star extends from 0.35\,$M_{\rm \odot}$ to 0.53\,$M_{\rm \odot}$.

However, inflation effects have been found in the case of hot Jupiter exoplanets \citep[e.g.][]{udalski} and also in the MS\,+\,BD binary CoRoT-15b \citep{bouchy}. 
As the companion is exposed to intense radiation of a luminous hot star at a distance of only
0.56\,$R_{\odot}$, this effect cannot be neglected and would result in an underestimation of the radius, if compared to non-irradiated models \citep{baraffe}.
The maximum inflation effect can be estimated from the comparison of our solutions to the theoretical mass-radius relations shown in Fig.~\ref{bd}. This figure shows that an inflation of more than about 10\,\% can be excluded, because otherwise none of the theoretical mass-radius relations would match the measured one, even if the star were as old as 10~Gyrs.

The mass-radius relation for the companion would be in perfect agreement with the light curve solution for a companion with a mass of 0.069\,$M_{\rm \odot}$, a radius of 0.086\,$R_{\rm \odot}$, and an age of $\sim$ 5-10~Gyrs, if we assume an inflation of 6-11\,\%. The corresponding mass of the sdB is 0.47\,$M_{\rm \odot}$, exactly the canonical sdB mass, which we therefore adopt for the sdB throughout the rest of the paper. A similar result was found in the analysis of the sdB\,+\,BD binary J1622 \citep{vs:2014_I}, which has a comparable period and parameters.

The companion has a mass below the limit for hydrogen-burning and thus appear to be a brown dwarf --  the third confirmed around a hot subdwarf star.

%

\section{Summary and conclusions}
We performed an analysis of the spectrum and light curve of V2008-1753 and find that this eclipsing binary consists of a pulsating sdB with a brown dwarf companion. This is the first system of this kind ever found. Similar to J1622, an inflation of the brown dwarf by more than about 10\,\% can be excluded.


V2008-1753 has the shortest period of all known HW~Vir systems and the second shortest period of any sdB binary discovered to date. Due to the small separation distance and high temperature of the sdB, the amplitude of the reflection effect is relatively large (more than 10\,\%).
Consequently, this system might provide the chance to detect and evaluate spectral features of the irradiated companion, similar to AA~Doradus and HW Vir \citep{aador,hw_vir}.
If the companion's spectral features are detected, the radial velocities of both components could be determined. We then could derive an unbiased mass ratio of the system and obtain a unique spectroscopic and photometric solution.

Higher quality and resolution spectra should also allow the detection of the Rossiter-McLaughlin (RM) effect \citep[cp.][]{nyvir}. This effect is due to the selective blocking of the light of the rotating star during an eclipse. The amplitude is mainly depending on radius ratio, the rotational velocity of the primary star, and the inclination of the system. As our system has a high inclination and a high radius ratio, the expected amplitude is quite high. From the RM effect it is, hence, possible to determine the rotational velocity independent from spectral line modelling. With the help of the rotational velocity we can check, if the sdB is synchronized, which should be the case due to synchronization theories, but is currently under debate \citep{vs:2014_I}.

The presence of stellar pulsations in the sdB offers further possibilities to characterize this system.   Once the pulsation frequencies are fully resolved and their modes identified with higher--precision data over a longer time base, the mass and other properties of the sdB can be constrained by asteroseismology.  These results can be compared to the light curve models.  To date, only NY\,Vir \citep{vangrootel} has offered this opportunity. However, no signature of the companion could be identified in the spectrum of this system, and the masses from the light curve analysis remain biased.

The combined presence of pulsations and eclipses, furthermore, offers the possibility of  eclipse mapping, as done for example for NY~Vir \citep{reed:2005, reed:2006}. Thereby, the eclipses can be used to determine pulsation modes, which are often difficult to uniquely identify. During the eclipse parts of the star are covered. Changing the amount and portion of regions of the star visible affects the pulsation amplitudes. The effect is changing for different pulsation modes, which can be identified in this way.

V2008-1753 has the potential to eventually replace NY\,Vir as the benchmark system for understanding sdB stars and their binary nature, if emission lines of the companion are detected. It would permit the direct comparison of independent techniques (namely light curve modelling, asteroseismology, spectroscopy, and radial velocity variations) used to derive the stellar parameters. Hence, the reliability of these methods and models could be checked as well as possible systematic errors of the derived parameters could be further investigated.  Most notably, the mass determined by asteroseismology could be checked at a high precision level, if the semi-amplitudes of the radial velocity curves could be determined for both components.

\begin{acknowledgements}
Based on observations with the SOAR Telescope. The SOAR Telescope is a joint project of: Conselho Nacional de Pesquisas Cientificas e Tecnologicas CNPq-Brazil, The University of North Carolina at Chapel Hill, Michigan State University, and the National Optical Astronomy Observatory. Fourier transformation results obtained with the software package FAMIAS developed in the framework of the FP6 European Coordination Action HELAS \url{(http://www.helas-eu.org/)}
\\
V.S. acknowledges funding by the Deutsches Zentrum f\"ur Luft- und Raumfahrt (grant 50 OR 1110).
  
\end{acknowledgements}


\bibliography{aabib}

\begin{thebibliography}{54}
\expandafter\ifx\csname natexlab\endcsname\relax\def\natexlab#1{#1}\fi

\bibitem[{{Almeida} {et~al.}(2012){Almeida}, {Jablonski}, {Tello}, \&
  {Rodrigues}}]{almeida:2012}
{Almeida}, L.~A., {Jablonski}, F., {Tello}, J., \& {Rodrigues}, C.~V. 2012,
  \mnras, 423, 478

\bibitem[{{Baraffe} {et~al.}(2003){Baraffe}, {Chabrier}, {Barman}, {Allard}, \&
  {Hauschildt}}]{baraffe}
{Baraffe}, I., {Chabrier}, G., {Barman}, T.~S., {Allard}, F., \& {Hauschildt},
  P.~H. 2003, \aap, 402, 701

\bibitem[{{Barlow} {et~al.}(2013){Barlow}, {Kilkenny}, {Drechsel}, {Dunlap},
  {O'Donoghue}, {Geier}, {O'Steen}, {Clemens}, {LaCluyze}, {Reichart},
  {Haislip}, {Nysewander}, \& {Ivarsen}}]{barlow:2012}
{Barlow}, B.~N., {Kilkenny}, D., {Drechsel}, H., {et~al.} 2013, \mnras, 430, 22

\bibitem[{{Barlow} {et~al.}(2012){Barlow}, {Wade}, \& {Liss}}]{barlow:romer}
{Barlow}, B.~N., {Wade}, R.~A., \& {Liss}, S.~E. 2012, \apj, 753, 101

\bibitem[{{Bouchy} {et~al.}(2011){Bouchy}, {Deleuil}, {Guillot}, {Aigrain},
  {Carone}, {Cochran}, {Almenara}, {Alonso}, {Auvergne}, {Baglin}, {Barge},
  {Bonomo}, {Bord{\'e}}, {Csizmadia}, {de Bondt}, {Deeg}, {D{\'{\i}}az},
  {Dvorak}, {Endl}, {Erikson}, {Ferraz-Mello}, {Fridlund}, {Gandolfi},
  {Gazzano}, {Gibson}, {Gillon}, {Guenther}, {Hatzes}, {Havel}, {H{\'e}brard},
  {Jorda}, {L{\'e}ger}, {Lovis}, {Llebaria}, {Lammer}, {MacQueen}, {Mazeh},
  {Moutou}, {Ofir}, {Ollivier}, {Parviainen}, {P{\"a}tzold}, {Queloz}, {Rauer},
  {Rouan}, {Santerne}, {Schneider}, {Tingley}, \& {Wuchterl}}]{bouchy}
{Bouchy}, F., {Deleuil}, M., {Guillot}, T., {et~al.} 2011, \aap, 525, A68

\bibitem[{{Charpinet} {et~al.}(1996){Charpinet}, {Fontaine}, {Brassard}, \&
  {Dorman}}]{charpinet:1996}
{Charpinet}, S., {Fontaine}, G., {Brassard}, P., \& {Dorman}, B. 1996, \apjl,
  471, L103

\bibitem[{{Charpinet} {et~al.}(2000){Charpinet}, {Fontaine}, {Brassard}, \&
  {Dorman}}]{charpinet:2000}
{Charpinet}, S., {Fontaine}, G., {Brassard}, P., \& {Dorman}, B. 2000, \apjs,
  131, 223

\bibitem[{{Claret} \& {Bloemen}(2011)}]{claret}
{Claret}, A. \& {Bloemen}, S. 2011, \aap, 529, A75

\bibitem[{{Clausen} {et~al.}(2012){Clausen}, {Wade}, {Kopparapu}, \&
  {O'Shaughnessy}}]{clausen:2012}
{Clausen}, D., {Wade}, R.~A., {Kopparapu}, R.~K., \& {O'Shaughnessy}, R. 2012,
  \apj, 746, 186

\bibitem[{{Dorman} {et~al.}(1993){Dorman}, {Rood}, \&
  {O'Connell}}]{Dorman:1993}
{Dorman}, B., {Rood}, R.~T., \& {O'Connell}, R.~W. 1993, APJ, 419, 596

\bibitem[{{Drechsel} {et~al.}(1995){Drechsel}, {Haas}, {Lorenz}, \&
  {Gayler}}]{drechsel:1995}
{Drechsel}, H., {Haas}, S., {Lorenz}, R., \& {Gayler}, S. 1995, A{\rm \&}A,
  294, 723

\bibitem[{{Drechsel} {et~al.}(2001){Drechsel}, {Heber}, {Napiwotzki},
  {{\O}stensen}, {Solheim}, {Johannessen}, {Schuh}, {Deetjen}, \&
  {Zola}}]{drechsel:2001}
{Drechsel}, H., {Heber}, U., {Napiwotzki}, R., {et~al.} 2001, A{\rm \&}A, 379,
  893

\bibitem[{{For} {et~al.}(2010){For}, {Green}, {Fontaine}, {Drechsel}, {Shaw},
  {Dittmann}, {Fay}, {Francoeur}, {Laird}, {Moriyama}, {Morris},
  {Rodr{\'{\i}}guez-L{\'o}pez}, {Sierchio}, {Story}, {Strom}, {Wang}, {Adams},
  {Bolin}, {Eskew}, \& {Chayer}}]{for:2010}
{For}, B.-Q., {Green}, E.~M., {Fontaine}, G., {et~al.} 2010, APJ, 708, 253

\bibitem[{{Geier} {et~al.}(2011){Geier}, {Schaffenroth}, {Drechsel}, {Heber},
  {Kupfer}, {Tillich}, {{\O}stensen}, {Smolders}, {Degroote}, {Maxted},
  {Barlow}, {G{\"a}nsicke}, {Marsh}, \& {Napiwotzki}}]{geier}
{Geier}, S., {Schaffenroth}, V., {Drechsel}, H., {et~al.} 2011, APJ, 731, L22+

\bibitem[{{Green} {et~al.}(2003){Green}, {Fontaine}, {Reed}, {Callerame},
  {Seitenzahl}, {White}, {Hyde}, {{\O}stensen}, {Cordes}, {Brassard}, {Falter},
  {Jeffery}, {Dreizler}, {Schuh}, {Giovanni}, {Edelmann}, {Rigby}, \&
  {Bronowska}}]{green:2003}
{Green}, E.~M., {Fontaine}, G., {Reed}, M.~D., {et~al.} 2003, \apjl, 583, L31

\bibitem[{{Han} {et~al.}(2012){Han}, {Chen}, {Lei}, \&
  {Podsiadlowski}}]{han:2012}
{Han}, Z., {Chen}, X., {Lei}, Z., \& {Podsiadlowski}, P. 2012, in Astronomical
  Society of the Pacific Conference Series, Vol. 452, Fifth Meeting on Hot
  Subdwarf Stars and Related Objects, ed. D.~{Kilkenny}, C.~S. {Jeffery}, \&
  C.~{Koen}, 3

\bibitem[{{Han} {et~al.}(2003){Han}, {Podsiadlowski}, {Maxted}, \&
  {Marsh}}]{han:2003}
{Han}, Z., {Podsiadlowski}, P., {Maxted}, P.~F.~L., \& {Marsh}, T.~R. 2003,
  MNRAS, 341, 669

\bibitem[{{Han} {et~al.}(2002){Han}, {Podsiadlowski}, {Maxted}, {Marsh}, \&
  {Ivanova}}]{han:2002}
{Han}, Z., {Podsiadlowski}, P., {Maxted}, P.~F.~L., {Marsh}, T.~R., \&
  {Ivanova}, N. 2002, MNRAS, 336, 449

\bibitem[{{Heber}(2009)}]{heber:2009}
{Heber}, U. 2009, ARA{\rm \&}A, 47, 211

\bibitem[{{Heber} {et~al.}(2000){Heber}, {Reid}, \& {Werner}}]{heber:2000}
{Heber}, U., {Reid}, I.~N., \& {Werner}, K. 2000, \aap, 363, 198

\bibitem[{Hirsch(2009)}]{hirsch}
Hirsch, H. 2009, Phd thesis, Friedrich Alexander Universit\"at Erlangen
  N\"urnberg

\bibitem[{{Ivanova} {et~al.}(2013){Ivanova}, {Justham}, {Chen}, {De Marco},
  {Fryer}, {Gaburov}, {Ge}, {Glebbeek}, {Han}, {Li}, {Lu}, {Marsh},
  {Podsiadlowski}, {Potter}, {Soker}, {Taam}, {Tauris}, {van den Heuvel}, \&
  {Webbink}}]{ivanova}
{Ivanova}, N., {Justham}, S., {Chen}, X., {et~al.} 2013, \aapr, 21, 59

\bibitem[{{Kilkenny} {et~al.}(1997){Kilkenny}, {Koen}, {O'Donoghue}, \&
  {Stobie}}]{kilkenny:1997}
{Kilkenny}, D., {Koen}, C., {O'Donoghue}, D., \& {Stobie}, R.~S. 1997, \mnras,
  285, 640

\bibitem[{{Kilkenny} {et~al.}(1998){Kilkenny}, {O'Donoghue}, {Koen},
  {Lynas-Gray}, \& {van Wyk}}]{kilkenny:1998}
{Kilkenny}, D., {O'Donoghue}, D., {Koen}, C., {Lynas-Gray}, A.~E., \& {van
  Wyk}, F. 1998, \mnras, 296, 329

\bibitem[{{Klepp} \& {Rauch}(2011)}]{klepp:2011}
{Klepp}, S. \& {Rauch}, T. 2011, A\rm \&A, 531, L7+

\bibitem[{{Kupfer} {et~al.}(2015){Kupfer}, {Geier}, {Heber}, {{\O}stensen},
  {Barlow}, {Maxted}, {Heuser}, {Schaffenroth}, \&
  {G{\"a}nsicke}}]{kupfer:2015}
{Kupfer}, T., {Geier}, S., {Heber}, U., {et~al.} 2015, ArXiv e-prints

\bibitem[{{Lucy}(1967)}]{lucy:1967}
{Lucy}, L.~B. 1967, Zeitschrift f\"ur Astrophysik, 65, 89

\bibitem[{{Markwardt}(2009)}]{markwardt:2009}
{Markwardt}, C.~B. 2009, in Astronomical Society of the Pacific Conference
  Series, Vol. 411, Astronomical Data Analysis Software and Systems XVIII, ed.
  D.~A. {Bohlender}, D.~{Durand}, \& P.~{Dowler}, 251

\bibitem[{{Maxted} {et~al.}(2002){Maxted}, {Marsh}, {Heber}, {Morales-Rueda},
  {North}, \& {Lawson}}]{maxted:2002}
{Maxted}, P.~F.~L., {Marsh}, T.~R., {Heber}, U., {et~al.} 2002, MNRAS, 333, 231

\bibitem[{{Maxted} {et~al.}(2006){Maxted}, {Napiwotzki}, {Dobbie}, \&
  {Burleigh}}]{nature:maxted}
{Maxted}, P.~F.~L., {Napiwotzki}, R., {Dobbie}, P.~D., \& {Burleigh}, M.~R.
  2006, \nat, 442, 543

\bibitem[{{Napiwotzki} {et~al.}(2004){Napiwotzki}, {Karl}, {Lisker}, {Heber},
  {Christlieb}, {Reimers}, {Nelemans}, \& {Homeier}}]{napi}
{Napiwotzki}, R., {Karl}, C.~A., {Lisker}, T., {et~al.} 2004, \apss, 291, 321

\bibitem[{{Nelemans} \& {Tauris}(1998)}]{nt}
{Nelemans}, G. \& {Tauris}, T.~M. 1998, \aap, 335, L85

\bibitem[{{{\O}stensen} {et~al.}(2010){{\O}stensen}, {Green}, {Bloemen},
  {Marsh}, {Laird}, {Morris}, {Moriyama}, {Oreiro}, {Reed}, {Kawaler}, {Aerts},
  {Vu{\v c}kovi{\'c}}, {Degroote}, {Telting}, {Kjeldsen}, {Gilliland},
  {Christensen-Dalsgaard}, {Borucki}, \& {Koch}}]{oestensen:2010}
{{\O}stensen}, R.~H., {Green}, E.~M., {Bloemen}, S., {et~al.} 2010, \mnras,
  408, L51

\bibitem[{{{\O}stensen} {et~al.}(2008){{\O}stensen}, {Oreiro}, {Hu},
  {Drechsel}, \& {Heber}}]{oestenson:2008}
{{\O}stensen}, R.~H., {Oreiro}, R., {Hu}, H., {Drechsel}, H., \& {Heber}, U.
  2008, in Astronomical Society of the Pacific Conference Series, Vol. 392, Hot
  Subdwarf Stars and Related Objects, ed. {U.~Heber, C.~S.~Jeffery, \&
  R.~Napiwotzki}, 221--+

\bibitem[{{Reed} {et~al.}(2005){Reed}, {Brondel}, \& {Kawaler}}]{reed:2005}
{Reed}, M.~D., {Brondel}, B.~J., \& {Kawaler}, S.~D. 2005, \apj, 634, 602

\bibitem[{{Reed} \& {Whole Earth Telescope Xcov 21 and 23
  Collaborations}(2006)}]{reed:2006}
{Reed}, M.~D. \& {Whole Earth Telescope Xcov 21 and 23 Collaborations}. 2006,
  \memsai, 77, 417

\bibitem[{{Schaffenroth} {et~al.}(2014{\natexlab{a}}){Schaffenroth}, {Geier},
  {Barbu-Barna}, {Heber}, {Kupfer}, \& {Cordes}}]{tuc_schaff}
{Schaffenroth}, V., {Geier}, S., {Barbu-Barna}, I., {et~al.}
  2014{\natexlab{a}}, in Astronomical Society of the Pacific Conference Series,
  Vol. 481, Astronomical Society of the Pacific Conference Series, ed. V.~{van
  Grootel}, E.~{Green}, G.~{Fontaine}, \& S.~{Charpinet}, 253

\bibitem[{{Schaffenroth} {et~al.}(2013){Schaffenroth}, {Geier}, {Drechsel},
  {Heber}, {Wils}, {{\O}stensen}, {Maxted}, \& {di Scala}}]{vs}
{Schaffenroth}, V., {Geier}, S., {Drechsel}, H., {et~al.} 2013, \aap, 553, A18

\bibitem[{{Schaffenroth} {et~al.}(2014{\natexlab{b}}){Schaffenroth}, {Geier},
  {Heber}, {Kupfer}, {Ziegerer}, {Heuser}, {Classen}, \& {Cordes}}]{vs:2014_I}
{Schaffenroth}, V., {Geier}, S., {Heber}, U., {et~al.} 2014{\natexlab{b}},
  \aap, 564, A98

\bibitem[{{Schuh} {et~al.}(2006){Schuh}, {Huber}, {Dreizler}, {Heber},
  {O'Toole}, {Green}, \& {Fontaine}}]{schuh:2006}
{Schuh}, S., {Huber}, J., {Dreizler}, S., {et~al.} 2006, \aap, 445, L31

\bibitem[{{Soker}(1998)}]{soker}
{Soker}, N. 1998, \aj, 116, 1308

\bibitem[{{Stetson}(1987)}]{stetson:1987}
{Stetson}, P.~B. 1987, \pasp, 99, 191

\bibitem[{{Udalski} {et~al.}(2008){Udalski}, {Pont}, {Naef}, {Melo}, {Bouchy},
  {Santos}, {Moutou}, {D{\'{\i}}az}, {Gieren}, {Gillon}, {Hoyer}, {Mayor},
  {Mazeh}, {Minniti}, {Pietrzy{\'n}ski}, {Queloz}, {Ramirez}, {Ruiz},
  {Shporer}, {Tamuz}, {Udry}, {Zoccali}, {Kubiak}, {Szyma{\'n}ski},
  {Soszy{\'n}ski}, {Szewczyk}, {Ulaczyk}, \& {Wyrzykowski}}]{udalski}
{Udalski}, A., {Pont}, F., {Naef}, D., {et~al.} 2008, \aap, 482, 299

\bibitem[{{Van Grootel} {et~al.}(2013){Van Grootel}, {Charpinet}, {Brassard},
  {Fontaine}, \& {Green}}]{vangrootel}
{Van Grootel}, V., {Charpinet}, S., {Brassard}, P., {Fontaine}, G., \& {Green},
  E.~M. 2013, \aap, 553, A97

\bibitem[{{Van Noord} {et~al.}(2013){Van Noord}, {Molnar}, \&
  {Steenwyk}}]{van_noord}
{Van Noord}, D.~M., {Molnar}, L.~A., \& {Steenwyk}, S.~D. 2013, ArXiv e-prints

\bibitem[{{von Zeipel}(1924)}]{zeipel:1924}
{von Zeipel}, H. 1924, MNRAS, 84, 665

\bibitem[{{Vu{\v c}kovi{\'c}} {et~al.}(2007{\natexlab{a}}){Vu{\v c}kovi{\'c}},
  {Aerts}, {{\"O}stensen}, {Nelemans}, {Hu}, {Jeffery}, {Dhillon}, \&
  {Marsh}}]{vuckovic:2007}
{Vu{\v c}kovi{\'c}}, M., {Aerts}, C., {{\"O}stensen}, R., {et~al.}
  2007{\natexlab{a}}, \aap, 471, 605

\bibitem[{{Vu{\v c}kovi{\'c}} {et~al.}(2007{\natexlab{b}}){Vu{\v c}kovi{\'c}},
  {Aerts}, {{\"O}stensen}, {Nelemans}, {Hu}, {Jeffery}, {Dhillon}, \&
  {Marsh}}]{nyvir}
{Vu{\v c}kovi{\'c}}, M., {Aerts}, C., {{\"O}stensen}, R., {et~al.}
  2007{\natexlab{b}}, \aap, 471, 605

\bibitem[{{Vu{\v c}kovi{\'c}} {et~al.}(2014){Vu{\v c}kovi{\'c}}, {Bloemen}, \&
  {{\O}stensen}}]{hw_vir}
{Vu{\v c}kovi{\'c}}, M., {Bloemen}, S., \& {{\O}stensen}, R. 2014, in
  Astronomical Society of the Pacific Conference Series, Vol. 481, Astronomical
  Society of the Pacific Conference Series, ed. V.~{van Grootel}, E.~{Green},
  G.~{Fontaine}, \& S.~{Charpinet}, 259

\bibitem[{{Vu{\v c}kovi{\'c}} {et~al.}(2008){Vu{\v c}kovi{\'c}}, {{\O}stensen},
  {Bloemen}, {Decoster}, \& {Aerts}}]{aador}
{Vu{\v c}kovi{\'c}}, M., {{\O}stensen}, R., {Bloemen}, S., {Decoster}, I., \&
  {Aerts}, C. 2008, in Astronomical Society of the Pacific Conference Series,
  Vol. 392, Hot Subdwarf Stars and Related Objects, ed. U.~{Heber}, C.~S.
  {Jeffery}, \& R.~{Napiwotzki}, 199

\bibitem[{{Wilson} \& {Devinney}(1971)}]{wilson:1971}
{Wilson}, R.~E. \& {Devinney}, E.~J. 1971, APJ, 166, 605

\bibitem[{{Wood} \& {Saffer}(1999)}]{Wood:1999}
{Wood}, J.~H. \& {Saffer}, R. 1999, MNRAS, 305, 820

\bibitem[{{Zima}(2008)}]{zima}
{Zima}, W. 2008, Communications in Asteroseismology, 157, 387

\bibitem[{{Zorotovic} {et~al.}(2010){Zorotovic}, {Schreiber}, {G{\"a}nsicke},
  \& {Nebot G{\'o}mez-Mor{\'a}n}}]{zorotovic:2010}
{Zorotovic}, M., {Schreiber}, M.~R., {G{\"a}nsicke}, B.~T., \& {Nebot
  G{\'o}mez-Mor{\'a}n}, A. 2010, \aap, 520, A86

\end{thebibliography}
\bibliographystyle{aa}

\end{document}